\documentclass[pra,10pt,aps,twocolumn,floatfix,superscriptaddress]{revtex4-1}
\usepackage{amsmath,amssymb,graphicx,braket}
\usepackage{hyperref}
\usepackage{xcolor}
\newcommand{\ketbra}[2]{\ensuremath{\ket{#1}\!\bra{#2}}}
\newcommand{\et}[1]{{\color{black}#1}}
\newcommand{\jj}[1]{{\color{black}#1}}

\begin{document}

\title{Quantum control of \et{frequency tunable} transmon superconducting qubits}
\author{J. J. Garc\'ia-Ripoll}
\affiliation{Institute of Fundamental Physics IFF-CSIC, Calle Serrano 113b, Madrid 28006 Spain}
\author{A. Ruiz-Chamorro}
\affiliation{Institute of Fundamental Physics IFF-CSIC, Calle Serrano 113b, Madrid 28006 Spain}
\author{E. Torrontegui}
\affiliation{Institute of Fundamental Physics IFF-CSIC, Calle Serrano 113b, Madrid 28006 Spain}
\email{eriktorrontegui@gmail.com}

\begin{abstract}
In this work we analyze the implementation of a control-phase gate through the resonance between the $\ket{11}$ and $\ket{20}$ states of two statically coupled transmons. We find that there are many different controls for the transmon frequency that implement the same gate with fidelities around $99.8\%$ ($T_1\et{=T_2^{*}=17}$ $\mu$s) and $99.9\et{9}\%$ ($T_1=\et{T_2^{*}=300}$ $\mu$s) within a time that approaches the theoretical limit. All controls can be brought to this accuracy by calibrating the waiting time and the destination frequency near the $\ket{11}-\ket{20}$ resonance. However, some controls, such as those based on the theory of dynamical invariants, are particularly attractive due to reduced leakage, \et{robustness against decoherence,} and their limited bandwidth.
\end{abstract}

\maketitle

\section{Introduction}

Transmon qubits presently dominate the quantum computation and quantum simulation landscape. They are mildly anharmonic qubits, a fact that restricts speed of operations and the strength interactions that can be used in single and two-qubit gates. Within this platform, we find a great variety of two-qubit gates, which include gates assisted by microwave pulses\ \cite{Chow2011}, parametrically modulated couplers\ \cite{McKay2016,Ganzhorn2019}, parametrically modulated qubits\ \cite{Reagor2018}, gates implemented with tuneable-frequency qubit-qubit resonances\ \cite{Dicarlo2009,Barends2014,Rol2019} and gates implemented with tuneable couplings\ \cite{Arute2019}.

Out of this list, the last two paradigms include some of the experiments with greatest fidelities, including $99.\et{3}\%$ in the case of tuneable-frequency gates \et{\cite{Kelly2014}} and $99.41\%$ for tuneable couplers\ \cite{Arute2019}, values which slowly approach the $99.9(1)\%$ record fidelities of trapped ions. In this work we study the possibility of improving these metrics, optimizing superconducting qubit gates to reduce errors down to the $10^{-3}-10^{-4}$ range. This reduction would be a dramatic increase in quantum volume\ \cite{Moll2018}, increasing the power of NISQ computations\ \cite{Preskill2018}, and opening the door to scalable error correction and fault-tolerant quantum computation.

Our research focuses on the resonant CZ gate demonstrated by DiCarlo et al.\ \cite{Dicarlo2009}, and later on scaled up by Barends et al.\ \cite{Barends2014} to setups with up to 9 qubits. This gate uses qubits that are parked at different frequencies, $\omega_1 > \omega_2,$ so that under normal conditions their interaction is suppressed. To make a two-qubit gate, the frequency of the high-laying qubit $\omega_1$ is brought down to a resonant condition between the transmon states that have two excitations, $\ket{11}$ and $\ket{20}.$ An adiabatic or quasiadiabatic ramp\ \cite{Rol2019,Barends2014} guarantees that the transmons are returned to their original conditions, with eigenstates suffering only phase shifts
\begin{equation}
  \label{eq:two-qubit-gate}
  \ket{ss'} \to \exp(-i\phi_s -i\phi_{s'}-i\phi_{11}\ketbra{11}{11}) \ket{ss'}.
\end{equation}

Our study focuses on different choices for ramping down the frequency of the control qubit $\omega_1(t).$ We will show that, provided that the ramps are slower than the anharmonicity, errors can be brought below $10^{-4}$ by tuning the waiting time and the distance from perfect resonance. Moreover, we engineer controls based on variational methods that are bandwidth limited, demand a smoother change in the flux applied to the qubit and minimize leakage errors $10^{-2}$ times below quasiadiabatic protocols. Finally, our research shows that using quasiadiabatic controls does not improve the resilience against spontaneous emission errors \et{and dephasing}.

This work shows that there is great potential for implementing high-fidelity quantum gates in existing setups\ \cite{Rol2019,Andersen2019}, with speeds that are competitive, with little to no changes to the setups. This should help improving the quality of ongoing applications of this gate, as well as inspire similar studies for other gate paradigms\ \cite{Arute2019}.

The paper is structured as follows. In Sects.\ \ref{sec:transmon} and\ \ref{sec:coupling} we introduce the quantum description of one and two coupled transmon qubits. In Sect.\ \ref{sec:cz-protocol} we explain how the energy level structure of the transmons supports a phase or CZ gate, by bringing the qubits close to the $\ket{11}-\ket{20}$ resonance. Section\ \ref{sec:control} introduces three approaches to the design of the qubit ramp $\omega^{(1)}_{01}(t)$ using fast-quasiadiabatic techniques (Sects.\ \ref{sec:faquad} and\ \ref{sec:slepian}), the invariants method (Sect.\ \ref{sec:invariants}) and a variational approximation to the transmon dynamics (Sect.\ \ref{sec:variational}). In Sect.\ \ref{sec:performance} we study the performance of these protocols and variations thereof. In Sect.\ \ref{sec:cz-simple} we show that just ramping down and up the frequency of the transmon produces rather large errors, all of which can be corrected by (i) slowing the ramp, (ii) tuning the destination frequency and (iii) the waiting time at the middle of the ramp. Section\ \ref{sec:cz-optimization} illustrates how these simple tweaks can bring the errors down to $10^{-6}$ within realistic times for an ideal qubit. Moreover \et{ in Sect.\ \ref{sect:decoherence}}, even for moderate qubit lifetimes, of $T_1=1\et{7}$ or $\et{300}$ $\mu$s, gate errors of $2\times 10^{-3}$ and $\et{1}\times 10^{-4}$ are feasible. Section\ \ref{sect:distortions} analyzes the performance of the different controls with respect to pulse bandwidth and distortions.

\section{Transmon model}

\subsection{Bare transmon}
\label{sec:transmon}

Our starting point is the standard transmon qubit model\ \cite{Koch2007}, a circuit that consists on a large capacitor that shunts a nonlinear inductance, which is implemented by a Josephson junction or a SQUID. In the number-phase representation, the Hamiltonian for this circuit reads
\begin{equation}
  \label{eq:transmon}
  \hat H_T = 4 E_C \hat{n}^2 - E_J \cos(\hat\varphi),
\end{equation}
with canonical operators $[\hat{n},\exp(i\hat{\varphi})]=\exp(i).$ The bare transmon Hamiltonian can be approximately solved in the number basis, using states $\hat{n}\ket{m}=m\ket{m},$ for $m\in\mathbb{Z},$ with the representation $\exp(i\hat{\varphi}) = \sum_{m}\ket{m}\bra{m+1},$ and a moderate cut-off $|m|\leq 10-20.$

In the limit $E_J/E_C\gtrsim 50,$ the transmon behaves as a weakly nonlinear harmonic oscillator
\begin{equation}
  \label{eq:trans_harm}
  \hat H_T \simeq 4 E_C \hat{n}^2 + \frac{1}{2}E_J\hat\varphi^2 - \frac{1}{24}E_J\hat\varphi^4,
\end{equation}
and can be solved analytically\ \cite{Koch2007}. Identifying $4E_C\sim \hbar^2/2m$ and $E_J\sim m \omega^2,$ the model reads
\begin{equation}
  \label{eq:weak_ho}
  \hat H = \hbar\omega_{01} \hat{a}^\dagger \hat{a} + \et{\frac{\hbar\alpha}{2}} \hat{a}^{\dagger\,2}\hat{a}^2.
\end{equation}
The frequency $\hbar\omega_{01} \simeq \sqrt{8E_CE_J}$ denotes the splitting between the two lowest energy states, $\ket{0}$ and $\ket{1}\simeq \hat{a}^\dagger\ket{0},$ which we use to encode a qubit. \et{The anharmonicity $h\alpha =-E_C$} is small but allows us to detune all higher energy states, $\ket{2},\ket{3}\ldots$ Note that the Fock operators are defined in terms of phase and number
\begin{equation}
  \label{eq:fock-operators}
  \hat{\varphi} = \left( \frac{8E_C}{E_J} \right)^{1/4}\frac{(\hat{a}+\hat{a}^\dagger)}{\sqrt{2}},\quad \hat{n} =i\left( \frac{E_J}{8 E_C} \right)^{1/4} \frac{(\hat{a}^\dagger-\hat{a})}{\sqrt{2}},\notag
\end{equation}
but have an implicit dependency on the transmon parameters.

In this work we are concerned with processes where we tune the transmon gap $\omega_{01}$ by manipulating the Josephson inductance $E_J.$  This tuning is facilitated by replacing the Josephson junction in the transmon with a SQUID: the magnetic flux that threads this loop determines its effective inductance $E_J(\phi)\sim E_J(0)\cos(2\phi/\Phi_0)$ and the properties of the qubit. Changing $E_J$ is equivalent to \textit{squeezing} the harmonic oscillator, an unitary process that can introduce decoherence, through leakage---transmon states $\ket{0},\ket{1}$ of the computational basis are mapped to excited states $\ket{2},\ket{3}\ldots$---or unwanted transitions between the computational states. One goal in the following sections will be to minimize the errors in these processes, preserving the transmon eigenstates to implement useful quantum operations.

\subsection{Coupled transmons}
\label{sec:coupling}

This work focuses on a setup with two capacitively coupled transmons that are detuned from each other. We want to design quantum controls where one of the qubit is ramped down in frequency, brought close to resonance, so as to implement a two-qubit quantum gate. The joint qubit model can be written as
\begin{equation}
  \label{eq:coupled-transmons}
 \hat H = \hat H_{T,a}(\phi) + \hat H_{T,b} + \frac{1}{2}\hbar g_C(\hat{n}_a-\hat{n}_b)^2 .
\end{equation}
where the coupling constant $g_C$ embodies the capacitive interaction, and $\hat H_{T,a}(\phi)$ and $\hat H_{T,b}$ are the Hamiltonians of the tunable and the parked qubits. \et{All the numerical simulations  \cite{SM} refer to these full Hamiltonians (\ref{eq:coupled-transmons}) and (\ref{eq:transmon}), however, in the following several transformations will be applied to derive different dependencies of the control $\phi$.} We can express Hamiltonian\ \eqref{eq:coupled-transmons} in the basis of eigenstates of the uncoupled problem. Since we are focused on manipulating qubits, we can focus on the subspace with up to two excitations. Assuming $\omega_a\geq \omega_b,$ this subspace is formed by states $\ket{00}, \ket{01}, \ket{10}, \ket{02}, \ket{11}$ and $\ket{20},$ in order of increasing energy. In this basis, the model is very well approximated by a Hamiltonian matrix of the form
\begin{equation}
  \label{eq:effective-H}
  \hat H_\mathrm{eff}  =\left( \begin{matrix}
      0 & 0 & 0 & 0 & 0 & 0 \\
      0 & \omega_b & J_{1} & 0 & 0 & 0 \\
      0 & J_{1} & \omega_a & 0 & 0 & 0 \\
      0 & 0 & 0 & 2 \omega_b-\alpha_b & J_{2}& 0 \\
      0 & 0 & 0 & J_{2} & \omega_a+\omega_b & J_{2} \\
      0 & 0 &0 & 0 & J_{2} & 2\omega_a - \alpha_a
    \end{matrix}\right).
\end{equation}
Here the frequency of the first qubit $\omega_a$ is the only tuneable parameter, depending on the control flux. The anharmonicities $\alpha_a$ and $\alpha_b$ are approximately constant, as they only depend on the capacitive energy. Finally, we have that $J_2\simeq \sqrt{2}J_1\propto g_C,$ but we cannot rely on this \et{when simulating the full dynamics} if we want to have accurate gates with precisions below $1\%.$

In the simulations that follow, without loss of generality, we will use the parameters from Ref.\ \cite{Rol2019}. This implies qubits with parameters
\begin{align}
  \label{eq:parameters}
 &\omega_a = 2\pi\times 6.91\mbox{GHz},\; &\alpha_a =-2\pi\times0.331\mbox{ GHz},\\
 &\omega_b = 2\pi\times 5.69\mbox{GHz},\; &\alpha_b =-2\pi\times0.300\mbox{ GHz},\notag\\
 &J_1 = 2\pi\times 14.3\mbox{ MHz},\notag,\; &J_2 =2\pi\times 20.2\mbox{ MHz}.
\end{align}

\subsection{Resonant CZ gate}
\label{sec:cz-protocol}

\begin{figure}[t]
  \centering
  \includegraphics[width=0.9\linewidth]{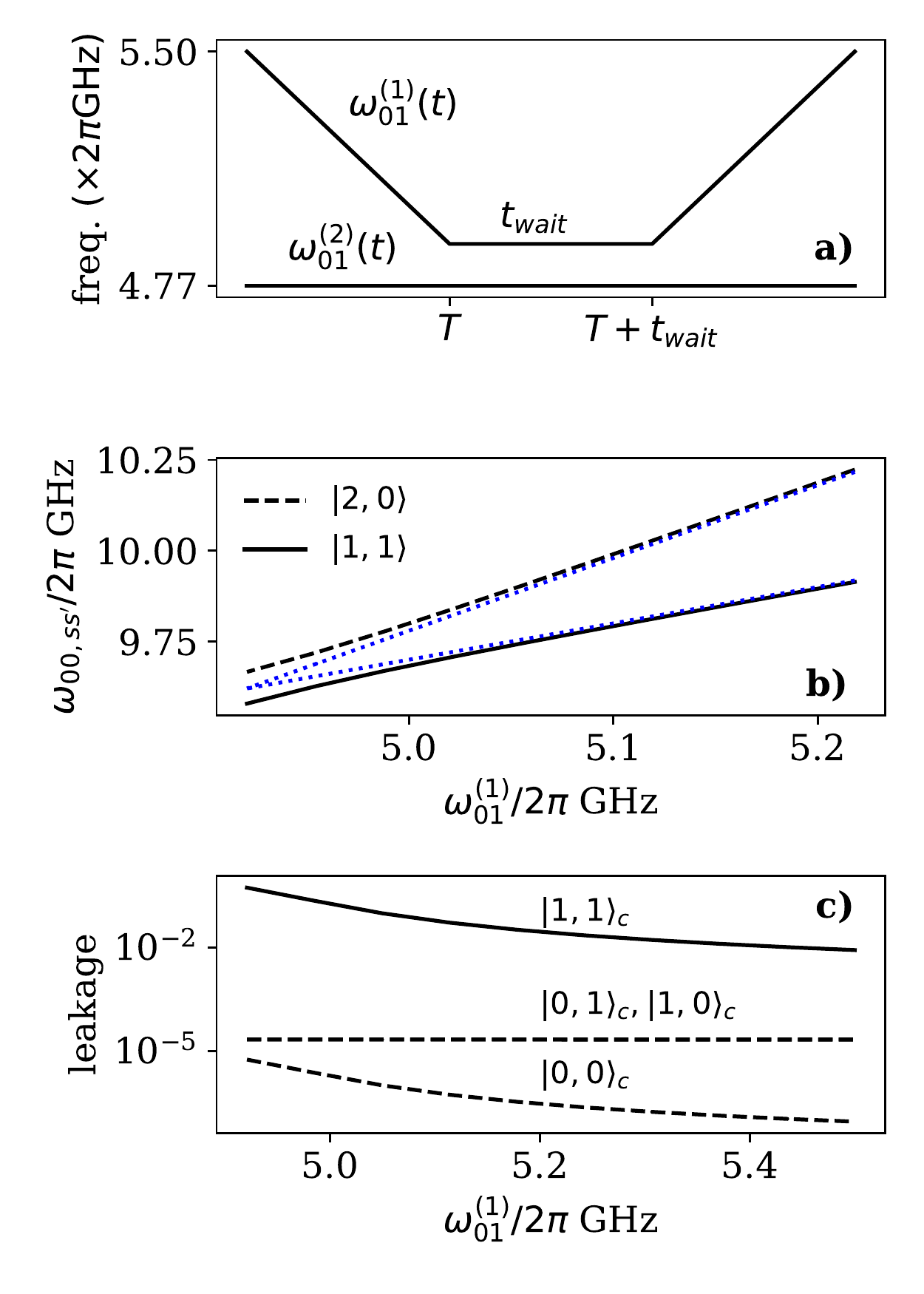}
  \caption{Implementation of a CZ gate. (a) Gate protocol: of two coupled transmon qubits---qubits Q0 and Q1 from Ref.\ \cite{Barends2014}---, one qubit is ramped down to a frequency where the states $\ket{11}$ and $\ket{20}$ of the transmon are degenerate. (b) Energy levels for two coupled Xmons\ \cite{Barends2014}, as we ramp the energy of one of the qubits. We plot the bare states $\ket{11}$ and $\ket{20}$ (dashed), together with the eigenstates of the coupled qubits. (c) Leakage from the coupled eigenstates outside the qubit computational basis, for different values of the detuning of $\omega_{01}^{(1)}$ from $\omega_{01}^{(2)}=2\pi\times4.77$ GHz.}
  \label{fig:2qb-cz-protocol}
\end{figure}

Assuming that $\omega_b$ is the smallest frequency and that $\omega_a$ can be tuned, the effective Hamiltonian\ (\ref{eq:effective-H}) has two avoided crossings. One at $\omega_a=\omega_b$ enables coherent exchange of excitations between the $\ket{01}$ and $\ket{10}$ qubit states. The second crossing, sketched in Fig.\ \ref{fig:2qb-cz-protocol}b happens at $\omega_a = \omega_b+\alpha_a$ and is a result of the interaction between the qubit state $\ket{11}$ and a state $\ket{20}$ outside the computational basis.

We will use this second avoided crossing to model a controlled-Z gate demonstrated in various experiments with transmon qubits\ \cite{Dicarlo2009,Barends2014,Rol2019}. Following the literature, we regard the subspace $\{\ket{11},\ket{20}\}$ as an effective pseudospin
\begin{equation}
  \label{eq:pseudospin}
  \hat H_{11,20}=\frac{1}{2}\delta \hat{\sigma}_z + J_2 \hat\sigma_x + \mathcal{O}\left(\frac{J_2^2}{\alpha_a+\alpha_b}\right)
\end{equation}
where we have full control of the longitudinal magnetic field $\delta(\phi) = \omega_a(\phi)-\omega_b-\alpha_a,$ with fixed transverse field $J_2.$ We will control $\delta(\phi)$ following the protocol from Fig.\ \ref{fig:2qb-cz-protocol}, bringing the qubits in and out of resonance. In the adiabatic limit, where $\omega_a$ changes much slower than the gap $J_2,$ Landau-Zener excitations are prevented and the ramp implements a phase gate
\begin{equation}
\label{eq:phase-gate}
  \hat U = \exp\left( i\phi_0 + i \phi_1 \hat\sigma^z_a + i \phi_2\hat\sigma^z_b + i\phi_{12}\hat\sigma^z_a\hat\sigma^z_b \right),
\end{equation}
which becomes a universal CZ operation for $\phi_{12}=\pi/4$. Here $\hat\sigma_{a,b}^z$ corresponds to the Pauli matrices acting on the $\ket{0}$ and $\ket{1}$ states of qubits $a$ and $b$ respectively.

It has been argued theoretically\ \cite{Martinis2014} and demonstrated experimentally\ \cite{Barends2014,Rol2019} that one needs not be perfectly adiabatic to implement this gate. The goal of the following sections is to provide different protocols for controlling the gate operation (\ref{eq:phase-gate})---i.e. the design of $\phi$ and therefore $\delta(\phi)$---, understanding sources of errors and the performance limits of the gate under realistic operations---e.g. limited bandwidth in the controls.

\section{Control theory}
\label{sec:control}

As mentioned above, a perfectly adiabatic gate can be a prohibitively demand for a realistic NISQ device. Fortunately there are many control designs that are robust and which allow us to implement the CZ gate in a time that approaches the ideal limit $\pi/J_2$ of instantaneous quenches---i.e. $T=0$ in Fig.\ \ref{fig:2qb-cz-protocol}b. Alternatively to standard optimal control theory\ \cite{Egger2014, Goerz2017, Basilewitsch2018}, we will achieve this limit making use of semi-analytic controls, presented below, that allow us to identify and correct several sources of error when designing the drivings. These controls are divided into two families. The FAQUAD and Slepian pulses aim at preserving the instantaneous eigenstates of the problem, minimizing the non-adiabatic corrections. The invariants and variational methods, on the other hand, aim at producing the right final state, allowing for high-order excitations that are self-corrected
at the end of the process.

\subsection{Generalized FAQUAD}
\label{sec:faquad}

The \textit{fast quasiadiabatic dynamics}\ \cite{Martinez-Garaot2015} method, is a technique that aims at preserving the adiabatic condition locally in time, to create fast and robust controls. We have extended this technique to consider excited states and problems with accidental degeneracies. Let us assume that we have a controlled Hamiltonian
\begin{equation}
  \hat H=\hat H_0+\epsilon(t)\hat H_1.
\end{equation}
We wish to engineer a quasiadiabatic passage $\epsilon(t)$ that preserves a subset of eigenstates $N=\{\ket{\psi_n(\epsilon)}\}.$ We will construct a larger set $\bar{N}$ that includes $N$ and all states that are spectral neighbors along the evolution---i.e. all states with energies immediately above $E_{n_{max}}$ or below $E_{n_{min}}$ those of $N$,  as well as all eigenstates in between---and which are potentially connected via $\hat H_0$ or $\hat H_1.$

Using these definitions, we now introduce an adiabaticiy parameter
\begin{equation}
  \mu(t) = \max_{r,n\in\bar{N}} \hbar\left| \frac{\braket{\psi_n(t) |\partial_t \psi_r(t)}}{E_r(t)-E_n(t)} \right|.
\end{equation}
This value estimates the rate of transition from $N$ to all other states. Imposing a small and constant transition rate $\mu(t)=c\ll 1$ we delocalize the transition probability along the whole interval and creates an equation for
the control $\epsilon(t)=\tilde\epsilon(t/T),$
\begin{equation}
  \label{eq:faquad}
  \tilde\epsilon(s) = \pm \frac{c}{\hbar}\int_{0}^{s=t/T}\max_{n,r\in \bar{N}}\left| \frac{E_n-E_r}{\braket{\psi_n|\partial_{\tilde\epsilon}\psi_r}} \right|\mathrm{d}s,
\end{equation}
that leads to the same control profile for any $T$ value. \et{For the design of the control (\ref{eq:faquad}), we consider the six levels containing zero, one, and two simultaneous excitations as they capture all the unitary dynamics of the two coupled transmons.}

\subsection{Slepian pulses}
\label{sec:slepian}

Martinis and Geller\ \cite{Martinis2014} have provided an alternative derivation of fast quasiadiabatic protocols that focus on the shape of the control, providing conditions to reduce the non-adiabatic corrections. Essentially, the control works with the pseudospin model\ \eqref{eq:pseudospin}, introducing the instantaneous angle
\begin{equation}
  \theta(t) = \arctan(2J_2/\delta(t)).
\end{equation}
The bandwidth limited controls assume a ramp from $\theta_i=\theta(0)=\theta(2T)$ to $\theta_f=\theta(T)$ and back, with no waiting time $t_\mathrm{wait}=0.$ The controls are designed as
\begin{equation}
  \theta[s(t)] = \theta_i + \sum_{n=1}^N \lambda_n\left[ 1-\cos\left( \frac{2\pi n s(t)}{2T} \right) \right],
\end{equation}
where the proper time $s(t)$ is obtained by solving
\begin{equation}
  t = \int_0^s \sin[\theta(\tau)]\mathrm{d}\tau.
\end{equation}
In order to ensure the condition $\theta(T)=\theta_f,$ we have to impose
\begin{equation}
  \theta_f = \theta_i + \sum_{n\mbox{ odd}} 2\lambda_n,
\end{equation}
which leaves $N-1$ free parameters to optimize.

\subsection{Invariants}
\label{sec:invariants}

FAQUAD is an effective method to implement a diagonal transformation, but the restriction of preserving the instantaneous eigenstates limits the maximal speed. There is a broad family of shortcuts to adiabaticity\ \cite{Torrontegui2013, Guery-Odelin2019} that ignore this restriction. The method of \textit{scaling laws or invariants} relies on an operator $\hat{I}(t),$ that is preserved by the evolution\ \cite{Lewis1969}
\begin{equation}
\label{eq:inv}
  \frac{d\hat{I}}{dt} = \frac{\partial\hat{I}}{\partial{t}} + \frac{i}{\hbar} [\hat{I}, \hat H],
\end{equation}
and which has imposed common eigenstates with the Hamiltonian at the beginning and end of evolution $t=0$ and $t=T$
\begin{equation}
\label{eq:BC}
  [\hat H(0), \hat{I}(0)] = [\hat H(T), \hat{I}(T)] = 0.
\end{equation}
This property is enough to ensure that the eigenstates of the initial problem $\hat H(0)$ are mapped to the corresponding eigenstates of $\hat H(T)$.

\et{For the design of the control} we will use the invariants method as it was designed for the harmonic oscillator\ \cite{Chen2010, Torrontegui2012}, ignoring the weak nonlinearity of our transmon $\alpha$, see Eq. (\ref{eq:trans_harm}) or (\ref{eq:weak_ho}). Let us define $\omega(t)=\sqrt{8E_CE_J(t)}/\hbar$ as the instantaneous frequency of the transmon model. The invariant associated with the single transmon Hamiltonian (\ref{eq:trans_harm}) becomes\ \cite{Lewis1982}
\begin{equation}
  \label{eq:invariant}
  \hat{I}(t) = \frac{4E_C}{\hbar^2}\left[ \rho(t) \hat{n} - \frac{\hbar^2\dot\rho(t)}{8E_C} \hat{\varphi}\right]^2 + \frac{\hbar^2}{16E_C}\frac{c^2}{\rho^2(t)}\hat\varphi^2,
\end{equation}
where $c$ is an arbitrary constant that we take as the initial gap of the problem $c=\omega(0),$ for convenience, \et{and $\rho\equiv\rho(t)$ is a free function satisfying (\ref{eq:inv})}
\begin{equation}
  \label{eq:ermakov}
  \ddot\rho + \omega^2(t) \rho = \frac{\omega^2(0)}{\rho^3},
  \end{equation}
  \et{with the imposed boundary conditions (\ref{eq:BC})}
\begin{align}
\label{eq:BCrho}
  &\rho(0)=1,\;\rho(T)=\sqrt{\omega(T)/\omega(0)}=:\gamma, \notag\\
  &\dot\rho(0)=\dot\rho(T)=\ddot\rho(0)=\ddot\rho(T)=0.
\end{align}
Our goal is now to \et{inverse engineer $\omega(t)$ given an appropriate design of $\rho(t)$}
\begin{equation}
\label{eq:omega_inv}
  \omega^2(t) = \frac{\omega^2(0)}{\rho^4} - \frac{\ddot\rho}{\rho}.
\end{equation}
In our work we have adopted a polynomial ansatz that satisfies the boundary conditions (\ref{eq:BCrho}),
\begin{equation}
  \rho(t) = \gamma + (1-\gamma)(1-t/T)\sum_{n=0}^{n_{max}}c_n(t/T)^n,
\end{equation}
with $c_0=1,\; c_1=2,\; c_3=3$ and the condition $\sum_nc_n=0.$ Already the fourth-order solution $n_{max}=4,$ with no free parameters $c_n$, provides a very good control, but global searchers over various cost-functions---e.g. nonlinear energy, fidelities, leakage, etc---can also be implemented.

\subsection{Variational ansatz}
\label{sec:variational}

The variational method is an alternative technique, in which we approximate the evolution of a state by a manually crafted ansatz, and then design the control to ensure that the initial and final form of our ansatz match the preserved eigenstates \et{\cite{Anderson1983, Huang2020}}. In our particular model, we just aim at preserving the vacuum state,
\begin{equation}
\label{eq:ansatz}
  \phi(x;\sigma,\beta) = \frac{1}{\sqrt{\sqrt{\pi}\sigma}}\exp\left( -\frac{x^2}{2\sigma^2} - i\frac{\beta}{2}x \right).
\end{equation}
Using the Lagrangian associated with the Schr\"odinger equation
\begin{equation}
  \mathcal{L}[\psi] := \frac{1}{2} \braket{\psi|i\hbar\partial_t\psi} - \frac{1}{2}\braket{i\hbar\partial_t\psi|\psi} - \braket{\psi|\hat H(t)|\psi},
\end{equation}
we construct a new Lagrangian for the variational parameters \et{and the Hamiltonian (\ref{eq:transmon})} as $L(\sigma,\beta) = \mathcal{L}[\phi(x;\sigma,\beta)]$
\begin{equation}
 L(\sigma,\beta)  = -\frac{\hbar}{4}\dot\beta\sigma^2 + E_J e^{-\sigma^2/4} - 2\frac{E_C}{\sigma^2} - 2E_C\beta^2\sigma^2,
\end{equation}
and find the optimal approximation to the evolution using the Lagrange equations,
\begin{equation}
  \frac{d}{dt}\frac{\partial{L}}{\partial\dot\beta} = \frac{\partial{L}}{\partial\beta},\;  \frac{d}{dt}\frac{\partial{L}}{\partial\dot\sigma} = \frac{\partial{L}}{\partial\sigma}.
\end{equation}
The only relevant equation is that of the radius $\sigma\equiv\sigma(t)$
\begin{equation}
  \hbar^2\frac{\ddot\sigma}{2} + 8E_CE_J e^{-\sigma^2/4}\sigma = \frac{(8E_C)^2}{\sigma^3}.
\end{equation}
As before, we solve for the control and impose boundary conditions \et{so (\ref{eq:ansatz}) becomes an eigenstate at initial and final times}
\begin{align}
  \label{eq:variational}
  &E_J(t) = \frac{e^{\sigma^2/4}}{8E_C}\left[ \frac{(8E_C)^2}{\sigma^4}-\frac{\hbar^2\ddot\sigma}{\sigma} \right],\\
  &\sigma(0) = \left( \frac{8 E_C}{E_J(0)} \right)^{1/4},\; \sigma(T) = \left( \frac{8 E_C}{E_J(T)} \right)^{1/4},\notag\\
  &\dot\sigma(0)=\ddot\sigma(0)=\dot\sigma(T)=\ddot\sigma(T)=0.\notag
\end{align}
Note how in the linear limit, in which $e^{-\sigma^2/4}\simeq 1,$ this control is identical to\ \eqref{eq:omega_inv} with the identifications
\begin{align}
  &\omega(t)=\sqrt{8E_CE_J(t)}/\hbar, \nonumber\\
  &\sigma(t) = \sigma(0)\rho(t).
\end{align}

\subsection{Error quantification}

To analyze the performance of our controls, we will use two figures of merit. The first and simplest one will be the \textit{leakage} of the $d=4$ qubit states outside the computational
basis $\{\ket{00},\ket{10},\ket{01},\ket{11}\}$, which we label with indices $s=1,2,3,4.$ We define leakage as the averaged probability that those states leave the computational subspace
\begin{equation}
  L[\hat U] = \left| 1 - \frac{1}{d^2}\sum_{s,s'=1}^d \left| \braket{s|\hat U(T)|s'} \right|^2 \right|.
\end{equation}
This quantity is different from zero when, say, states such as $\ket{11}$ experience non-recoverable transitions to nearby states, such as $\ket{02}$ or $\ket{20}.$

The second figure of merit will be the average fidelity. As explained in Ref.\ \cite{Nielsen2002}, the average fidelity of a positive map $\mathcal{E}(\hat\rho)$ is a measure of how well quantum states are preserved by that channel
\begin{equation}
  \bar{\mathcal{F}}[\mathcal{E}] = \int \braket{\psi|\mathcal{E}(\ketbra{\psi}{\psi})|\psi}\mathrm\psi.
\end{equation}
Computing this quantity requires integrating over a Hilbert subspace of pure states $\ket{\psi}$ in the computational basis, using the uniform Haar measure. Instead of performing this integral, the average fidelity can be deduced from the entanglement fidelity\ \cite{Nielsen2002},
\begin{equation}
 \label{eq:avg-fidelity}
\bar{\mathcal{F}}[\mathcal{E}]= \frac{N\mathcal{F}_\text{e}[\mathcal{E}]+1}{N+1}.
\end{equation}
The entanglement fidelity is much easier to compute,
\begin{equation}
  \label{eq:ent-fidelity}
  \mathcal{F}_\text{e}[\mathcal{E}] := \braket{\phi|\left(\openone \otimes \mathcal{E}\right)(\ketbra{\phi}{\phi})|\phi}.
\end{equation}
because it is defined in terms of a single, maximally entangled state---for instance $\ket{\phi} = \sum_s \frac{1}{\sqrt{d}}\ket{s,s}.$

The average channel fidelity is a useful measure to compare the evolution of a controlled system $\hat U(T)$ with the ideal that we wish to implement $\hat U_\text{id}.$ To do this comparison, we compute the average fidelity over a positive map that does the real operation, followed by the inverse of the desired gate $\mathcal{E}_\text{comp}(\hat\rho) =\hat U_\text{id}^\dagger\hat U(T)\hat\rho\hat U^\dagger(T)\hat U_\text{id}.$ When $\hat U(T)$ and $\hat U_\text{id}$ coincide, the map is the identity and the fidelity is 1.

In actual simulations we tweak this approach, introducing an operation $\hat U_\text{loc}$ that eliminates all locally correctable phases. This way, we define the entanglement fidelity of our controlled unitary $\hat U(T)$ as
\begin{equation}
\mathcal{F}_e[\hat U_\text{id},\hat U(T)] :=  \left| \frac{1}{d}\sum_{s=1}^d \braket{s|\hat U_\text{id}^\dagger\hat U_\text{loc}^\dagger\hat U(T)|s}\right|^2,
\end{equation}
and use Eq.\ \eqref{eq:avg-fidelity} to deduce the average gate fidelity $\bar{\mathcal{F}}.$

In Sect.\ \ref{sect:decoherence} we study the implementation of a gate under realsitic dephasing and dissipation. In those cases the evolution of the system is given by a positive map, $\mathcal{E}_T(\hat\rho(0))=\hat\rho(T).$ Once more, we use the average fidelity to estimate how far this channel is from the desired two-qubit gate, up to local operations. The only difference is that now in the entanglement fidelity we have to compose the full non-unitary channel with the ideal gate and locally corrected phases, which gives the expression
\begin{align}
  \mathcal{F}_\text{e}[\hat U_\text{id},\mathcal{E}_T] =
  \frac{1}{d^2} \sum_{s,s'} \braket{s |\hat U_\text{id}^\dagger\hat U_\text{loc}^\dagger \mathcal{E}_T(\ketbra{s}{s'})\hat U_\text{loc}\hat U_\text{id} |s'}.
\end{align}
and use Eq.\ \eqref{eq:avg-fidelity} to deduce the average fidelity.

\section{Performance analysis}
\label{sec:performance}

\subsection{Ramping an isolated transmon}
\label{sec:ramping}

\begin{figure}[t]
  \centering
  \includegraphics[width=0.9\linewidth]{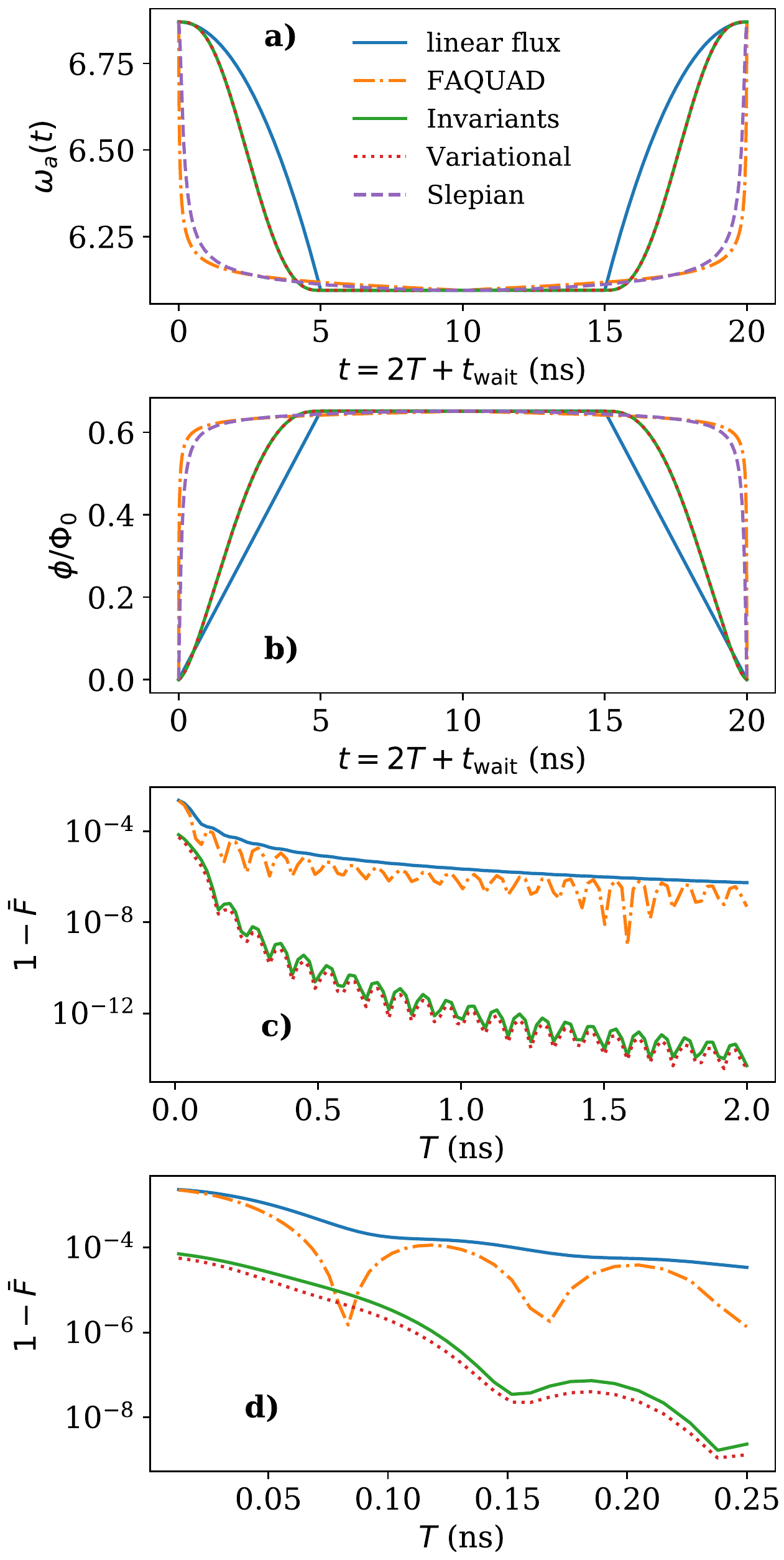}
  \caption{Ramp of an isolated transmon. (a) Possible controls for a total down-and-up ramp with $2T+t_\mathrm{wait}=20$ ns. \et{In this case $T=5$ ns and $t_\mathrm{wait}=10$ ns.} (b) Required change in the flux applied to the transmon to implement the control. (c) Average gate error for different lengths of the control, with $t_\mathrm{wait}=0.$ (d) Zoom in at extremely short controls. All simulations use the first qubit in\ \eqref{eq:parameters}.}
  \label{fig:1qb-ramp}
\end{figure}

As warmup problem we have studied how to change the gap of an isolated transmon, implementing the protocol from Fig.\ \ref{fig:2qb-cz-protocol}a without interactions. Figure\ \ref{fig:1qb-ramp}a illustrates the frequency change of the qubit for the controls from Sect.\ \ref{sec:control}. Note how the FAQUAD method accelerates in the regions of the passage that have a large gap, while it slows downs close to the crossing. The Slepian pulses from\ \cite{Martinis2014} find a similar behavior through a different reasoning.

Remember, however, that in order to tune the frequency of the transmon we have to thread a flux through its SQUID. The change in flux required to implement the controls are shown in Fig.\ \ref{fig:1qb-ramp}b. In solid blue line we draw a simple control that uses a linear ramp. The invariants and variational controls follow hardware friendly paths with vanishing slopes at the beginning and the end. Finally, both the FAQUAD and the Slepian controls exhibit a nasty behavior at these extremes: since $d\omega/dt$ is finite for these methods close to the sweet spot, it requires a diverging flux derivative to implement such pulses.

Figures\ \ref{fig:1qb-ramp}c and\ \ref{fig:1qb-ramp}d show the average fidelity of the down ramp (or the symmetric up ramp) for the different protocols, as a function of the ramp time $T$. % for $t_\mathrm{wait}=0.$ 
Remarkably, the linearly growing pulse exhibits as good a behavior as the quasiadiabatic methods, but all of them are well separated from the invariants and variational controls, which are the best performing methods.

Note how these controls provide errors below $10^{-6}$ for any ramp above $0.1$ ns, which is on the limit of the fastest ramps available in the laboratory. These two controls perform so well because they are essentially tracking the full dynamics of the zero and one excitation subspaces, which behave like the eigenstates of the harmonic oscillator. In particular, these protocols reproduce perfectly the squeezing of the oscillator and its eigenstates, down to very high precision. Interestingly, we have attempted to create optimal control pulses using parameterized methods and global optimizations---see App. I from Ref.\ \cite{Romero-Isart2007}---, but the fidelities were comparable at very large computational cost.

\begin{figure}[t]
  \centering
  \includegraphics[width=0.9\linewidth]{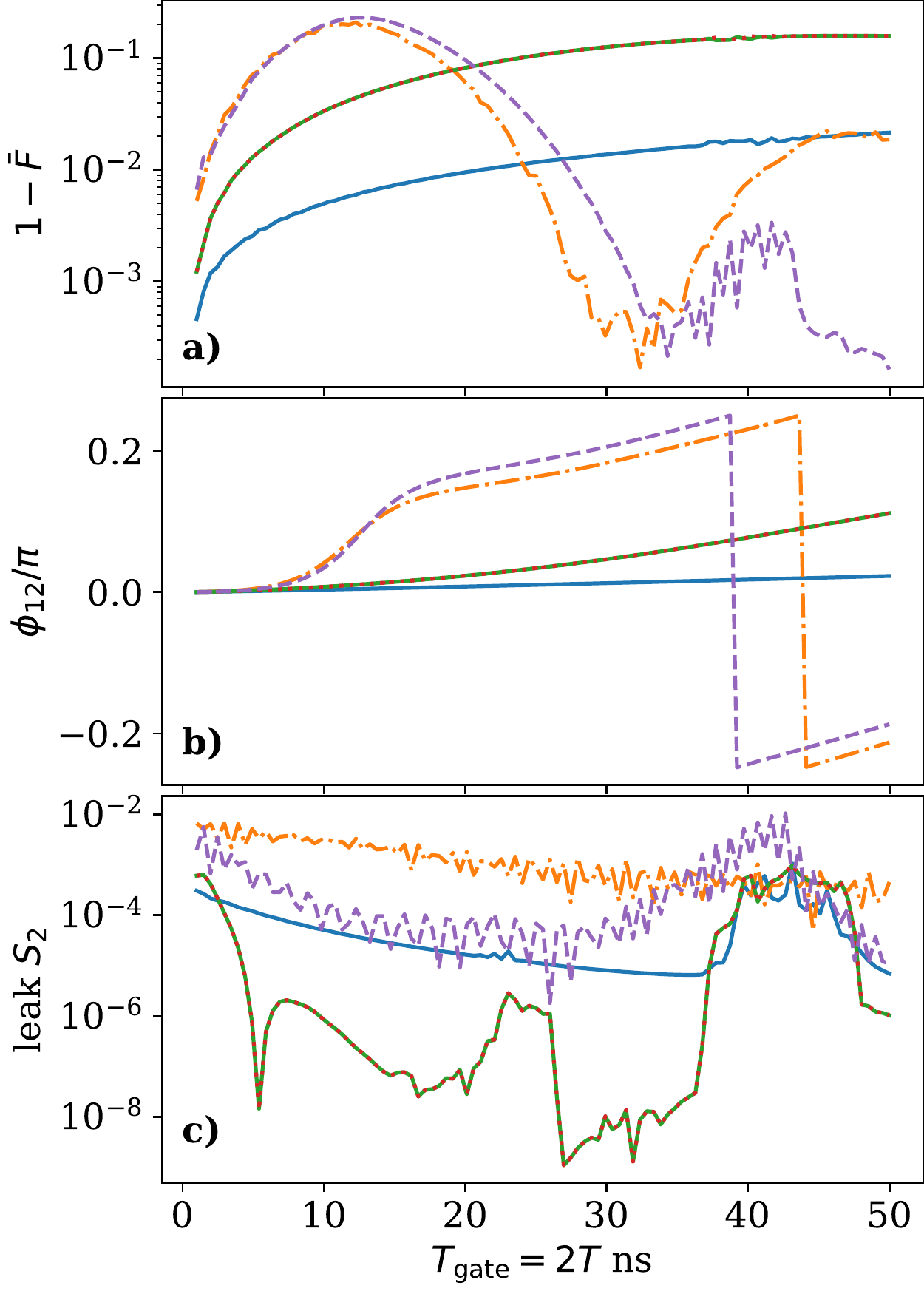}
  \caption{CZ gate by ramping down and up the qubit with $t_\mathrm{wait}=0.$ (a) Average gate infidelity from a generic phase gate. (b) Two-qubit phase acquired at the end. (c) Leakage of the evolved state $\hat U(2T)\ket{11}$ outside the $\{\ket{11},\ket{20}\}$ subspace. The line code follows Fig.\ \ref{fig:1qb-ramp}a.}
  \label{fig:2qb-cz-ramp}
\end{figure}

\subsection{CZ gates with simple ramps}
\label{sec:cz-simple}

We have studied the possibility of implementing the CZ gate using the protocol in Fig.\ \ref{fig:2qb-cz-protocol}b with $t_\mathrm{wait}=0.$ In this approach, the qubit is ramped down and up and we inspect the resulting operation. This choice is very natural for the FAQUAD and Slepian protocols which, as shown in Fig.\ \ref{fig:1qb-ramp}a have a built-in waiting time around the avoided crossing.

Figure\ \ref{fig:2qb-cz-ramp}a shows the average fidelity of the unitary operation acting in the qubit subspace, compared with the phase gate\ \eqref{eq:phase-gate} that approximates it the best. In this figure both the FAQUAD and the Slepian pulses achieve a reasonable accuracy, with an error below $0.1\%$ in a time around 30 ns, which is only slightly larger than the ideal limit $\pi/J_2.$ Out of these, the Slepian pulse even reaches the desired phase $\phi_{12}=\pi/4$ close to this fidelity, see Fig.\ \ref{fig:2qb-cz-ramp}b, while the FAQUAD protocol only achieves this phase in a region where the fidelity is bad again.

A na\"{\i}ve interpretation of these simulations would lead us to discard all protocols but the bandwidth limited controls\ \cite{Martinis2014}. However, if we investigate the errors further, we will find that they can be attributed to leakage from the $\ket{11}$ state into the $\ket{20}.$ Essentially, what is happening in all controls---including the FAQUAD and Slepian method---is that the two-qubit system approaches the resonance condition $\omega_a= \omega_b+\alpha$ a little faster than desired. This causes a Landau-Zener transition, with population that, once we ramp back, ends up in the $\ket{20}$ state.

\begin{figure}[t]
  \centering
  \includegraphics[width=0.8\linewidth]{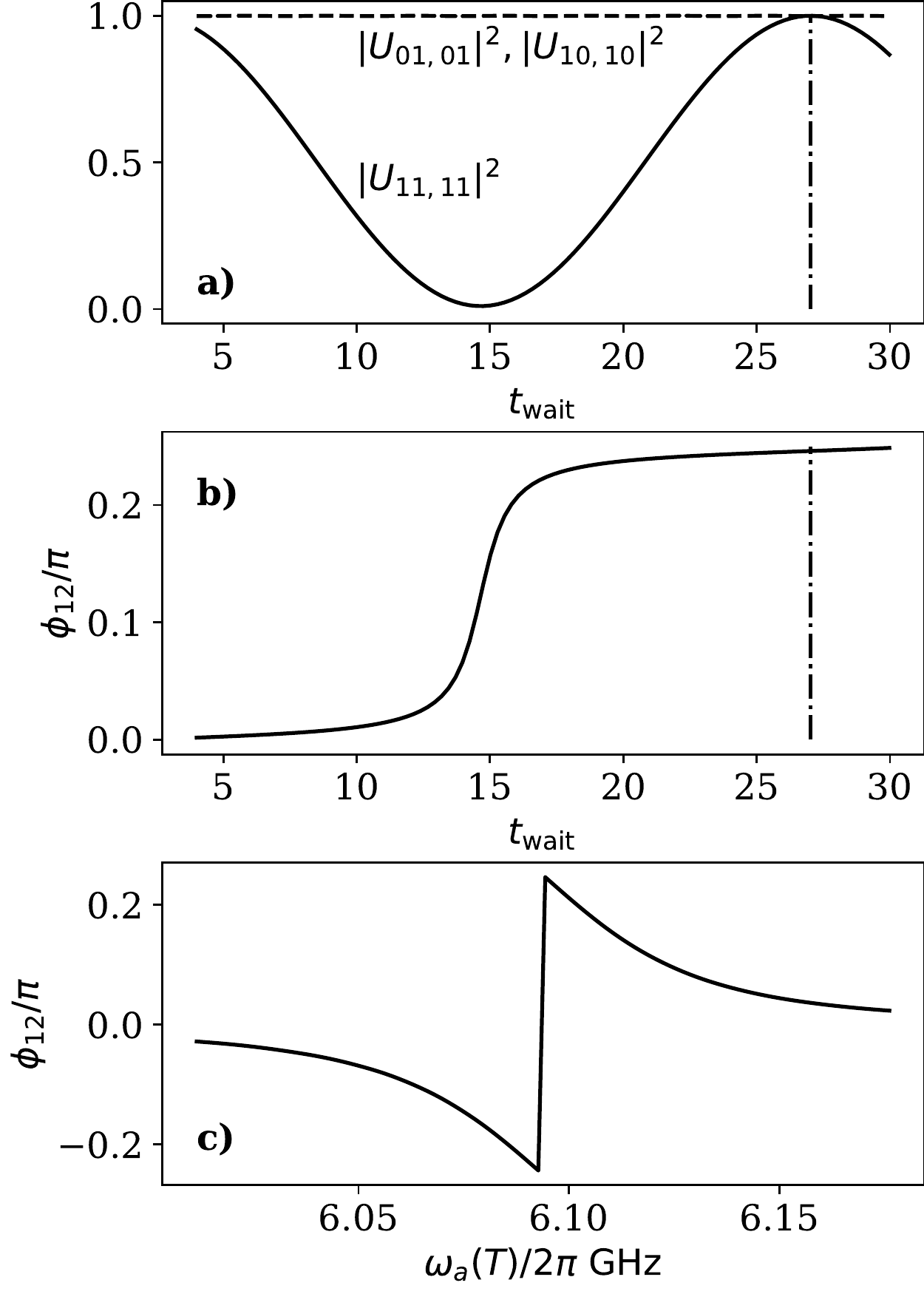}
  \caption{Recovery from non-adiabatic error for the invariant control with $T=2$ ns. (a) Transition probabilities as a function of the waiting time, for fixed ramp duration $T=2$ns. (b) Accumulated nonlinear phase $\phi_{12}$ as a function of the waiting time. (c) Accumulated phase for different final destination frequencies $\omega_a(T)$.}
  \label{fig:cz-wait-time}
\end{figure}

An even more careful inspection of the dynamics of the two-qubit system reveals that all the dynamics takes place within three separate subspaces, with different number of excitations $S_0:=\{\ket{00}\},$ $S_1:=\{\ket{01},\ket{10}\}$ and $S_2:=\{\ket{02},\ket{11},\ket{20}\}.$ Leakage outside $S_0$ and $S_1$ is negligible for $T\geq 0.5$ns, while leakage of $S_2$ is also small for some of our control protocols, as shown in Fig.\ \ref{fig:2qb-cz-ramp}c. Thanks to this, we can correct these errors, by just adding some wait time $t_\mathrm{wait},$ as shown below.

\subsection{CZ gate optimization}
\label{sec:cz-optimization}

\begin{figure}[t]
  \centering
  \includegraphics[width=0.9\linewidth]{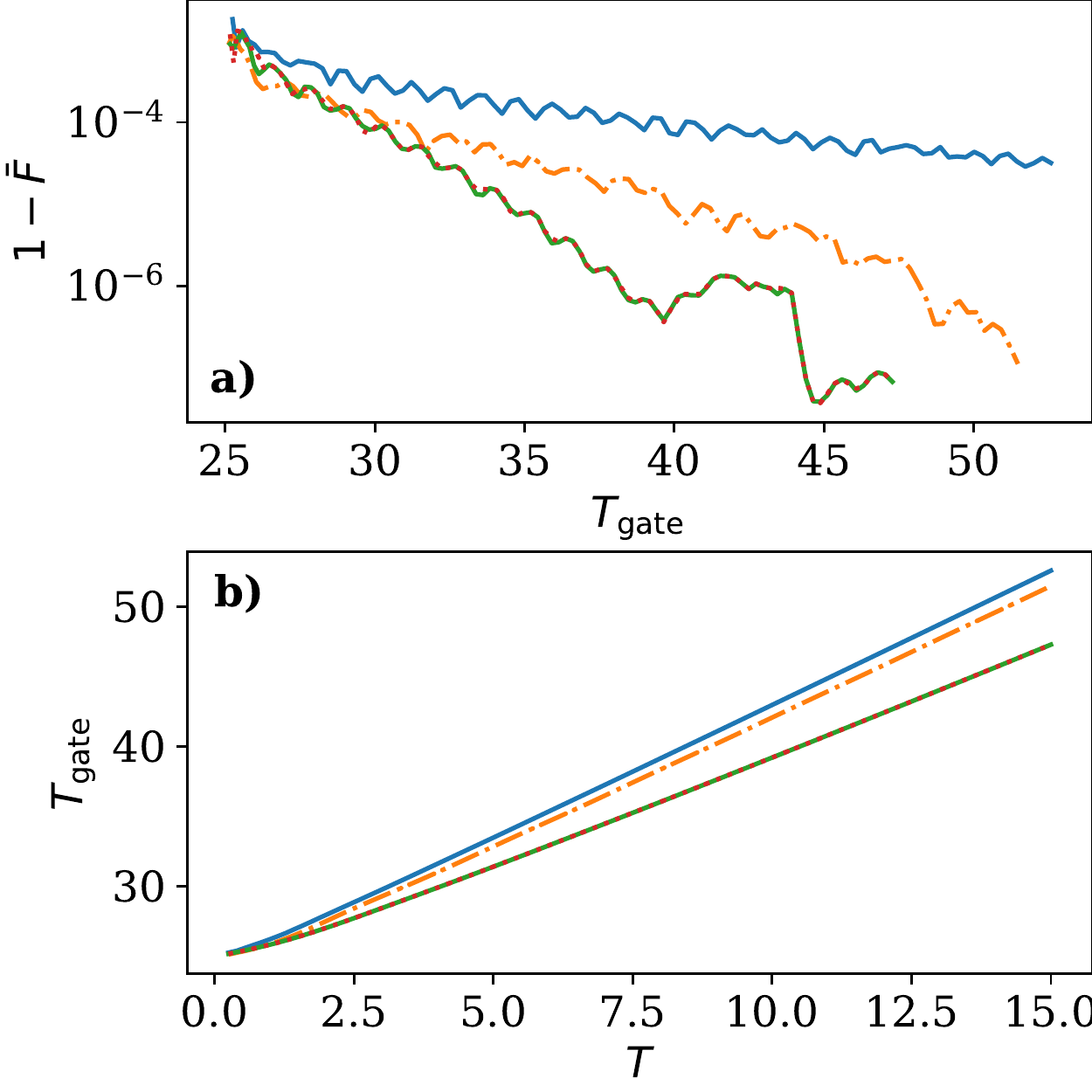}
  \caption{Optimized CZ gate. (a) Optimal fidelity vs. total gate time $T_\mathrm{gate}.$ (b) Total gate time for different ramp times.}
  \label{fig:optimal-cz}
\end{figure}

If we analyze the evolution of the qubit states, we find that all CZ controls suffer from the same errors: (i) when reaching the crossing point, some leakage from $\ket{11}$ to $\ket{20},\ket{02}$ states happens, (ii) the states $\ket{01}$ and $\ket{10}$ acquire some phase, making $\phi_{12}$ \et{deviate from $\pi/4,$} and (iii) there is some residual leakage states with higher number of excitations. Each source of error is best illustrated by each of the subfigures in Fig.\ \ref{fig:2qb-cz-ramp}, but they are all highly correctable.

As mentioned above, the dynamics takes place mostly in the zero to two excitation subspace. Moreover, the ramp-down and ramp-up operators are related $\hat U(2T,T)^T = \hat U(T,0),$ and they both have a simple structure when written in terms of the initial and final eigenstates
\begin{equation}
  \hat U(T,0) \simeq \left(
    \begin{matrix}
      1 & 0 & 0 & 0 & 0 & 0\\
      0 & e^{-i\xi_1 \et{T}} & 0 & 0 & 0 & 0\\
      0 & 0 & e^{-i\xi_2 \et{T}} & 0 & 0 & 0\\
      0 & 0 & 0 & e^{-i\xi_3\et{T}} & 0 & 0\\
      0 & 0 & 0 & 0 & \alpha  & \gamma \\
      0 & 0 & 0 & 0 & \beta & \delta
    \end{matrix}
  \right),
\end{equation}
As illustrated in this equation, states $\ket{00},\ket{01},\ket{10}$ and $\ket{02}$ are mostly mapped to eigenstates of the coupled system, modulo some phases. The last block is a $2\times2$ unitary operation that maps state $\ket{11}$ to a combination $\alpha\ket{-}+\beta\ket{+}$ of the pseudospin superposition $\ket{\pm}\propto\ket{11}\pm\ket{20}$. Since $\hat U(2T,0) =\hat U(T,0)^T\hat U(T,0),$ if we do not wait any time and simply ramp up, this state is mapped to $(\alpha^2 +\beta^2)\ket{11} + (\gamma\alpha + \beta\delta)\ket{20}$ at the end, which accounts for most errors in Fig.\ \ref{fig:2qb-cz-ramp}a.

This leakage is corrected by parking the qubits close to resonance for a certain time $t_\mathrm{wait}.$ The last $2\times2$ block in $\hat U(T,0)$ is an approximate unitary, which can be undone by waiting some time close to degeneracy, where states $\ket{\pm}$ freely evolve with different energies
\begin{align}
  &\hat U(T,0)^T e^{-i\hat H t} \hat U(T,0)\ket{11}  \\
  &\simeq (\alpha^2 e^{iJ_2t} + \beta^2 e^{-iJ_2t}) \ket{11} + (\gamma\alpha e^{iJ_2t}+\beta\delta e^{-iJ_2t})\ket{20}.\notag
\end{align}
Neglecting leakage into other states, we always find a time $e^{iJ_2t} = \beta/\alpha$ at which this state becomes identical to $\ket{11}$---the contribution of $\ket{20}$ cancels due to unitarity ($|\alpha|^2+|\beta|^2=1$) and we neglect leakage to other states---. Figure\ \ref{fig:cz-wait-time}a illustrates this for one particular control, the dynamical invariant method with a ramp-down and up time of $T=2$ ns. In this particular case, the initial leakage ($t_{wait}=0$) was about 1\%, but this leakage is corrected by waiting about 27 ns.

The condition of matching perfectly the population of the $\ket{11}$ state implies also a (-1) phase shift, caused by a $\pi$ rotation of the pseudospin. However, as seen in Fig.\ \ref{fig:cz-wait-time}b, the combined nonlinear phase still \et{deviate from $\phi_{12}=\pi/4,$} because of dynamical phases in the $\ket{01}, \ket{10}$ and $\ket{11}$ states. We correct these phases ramping down the qubit to a frequency that deviates slightly from the target value $\omega_a(T)=\omega_b+\alpha_a.$ As shown in Fig.\ \ref{fig:cz-wait-time}c, changes in the phase are linear with respect to this detuning, which will be of a few megahertz and within experimental reach.

With all these correction mechanisms---i.e. optimizing the unitary with respect to $t_\mathrm{wait}$ and the destination frequency  $\omega_a(T)$---we obtain at least two orders of magnitude increase in gate fidelity, as seen in Fig.\ \ref{fig:optimal-cz}a, irrespective of the control that is applied. Controls such as the invariants method perform extremely well, due to their capacity to address the oscillator squeezing and minimize leakage to states outside the computational basis, but a trivial linear ramp of the flux reaches gate fidelities above $99.9\%$  and $99.99\%$ for gate durations of 25 to 40 ns. \et{The remaining errors are due to leakage produced by unwanted transitions originated by the lack of adiabaticity (FAQUAD and Slepian) or energetic differences with respect to the uncoupled model for which the controls were designed (invariant and variational).}

\begin{figure}[t]
  \centering
  \includegraphics[width=0.9\linewidth]{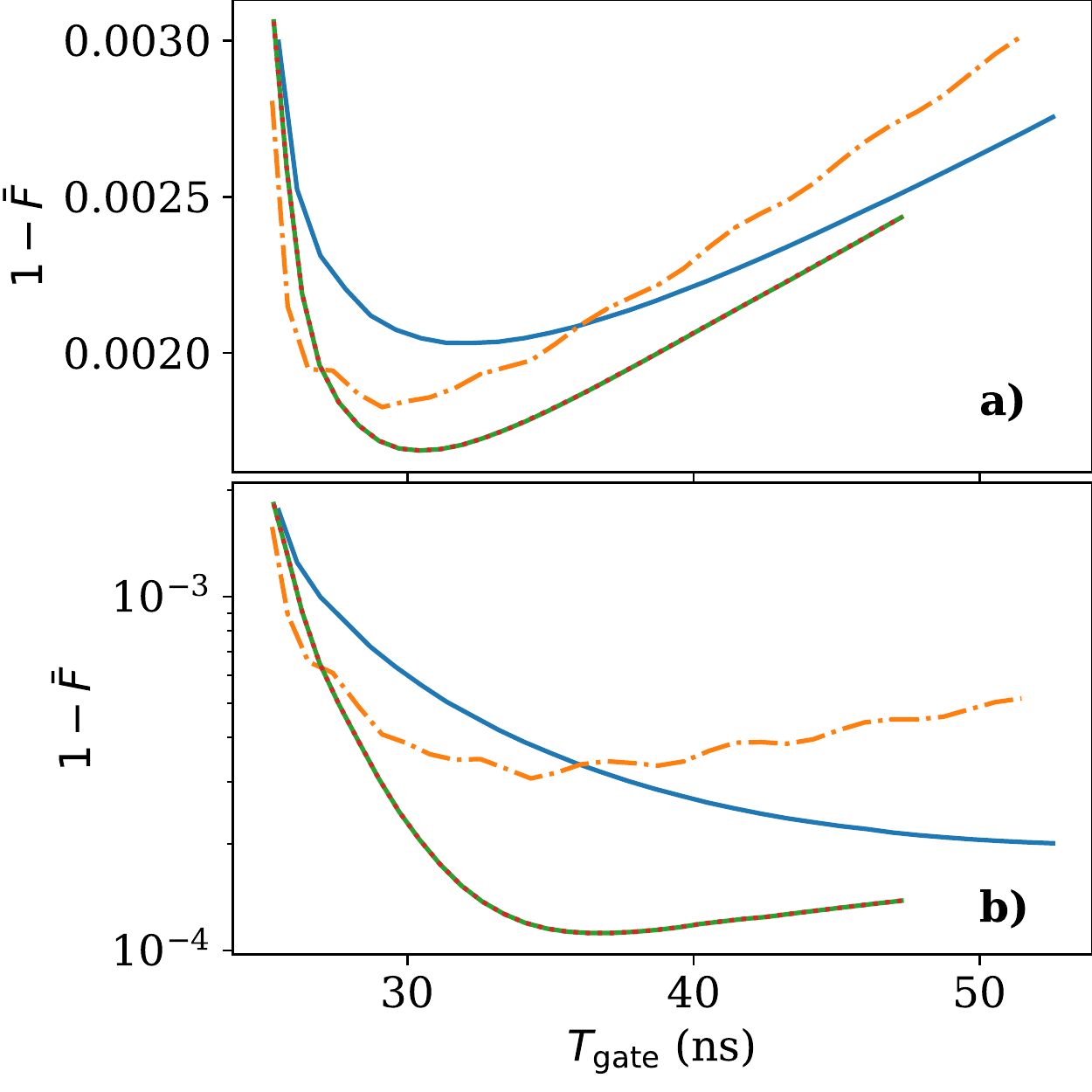}
  \caption{Optimized CZ gate for lossy qubits with (a) $T_1=\et{17}$ $\mu$s and (b) $T_1=\et{300}$ $\mu$s.}
  \label{fig:lossy-cz}
\end{figure}

\section{Imperfections}

\subsection{Decoherence}
\label{sect:decoherence}

\begin{figure}[t]
  \centering
  \includegraphics[width=0.9\linewidth]{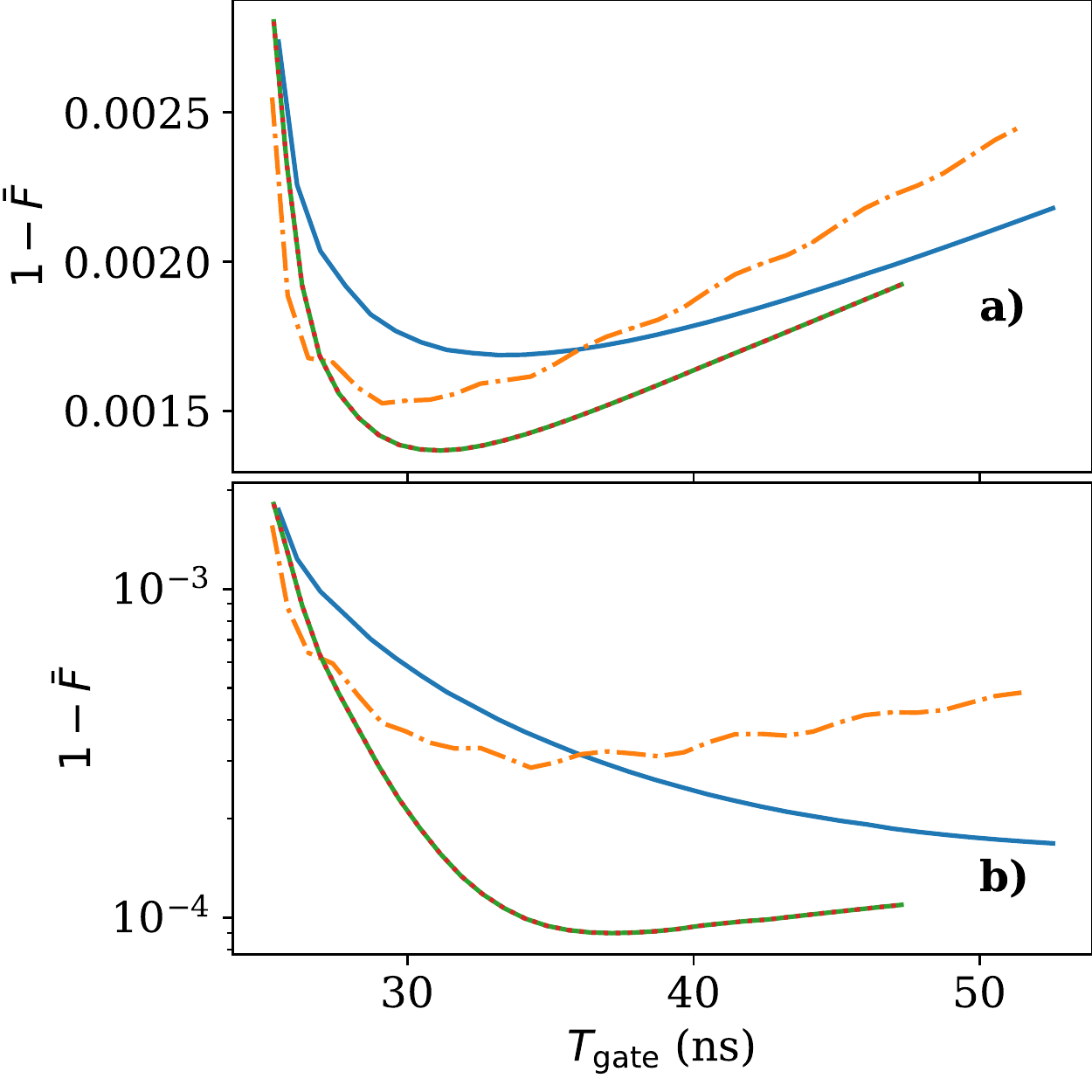}
  \caption{\et{Optimized CZ gate in the pressence of dephasing with (a) $T_2^{*}=17$ $\mu$s and (b) $T_2^{*}=300$ $\mu$s.}}
  \label{fig:dephased-cz}
\end{figure}
The same study can be done including losses and dephasing in the superconducting qubits. \jj{We model decoherence using a Lindblad master equation to describe the dynamics,
\begin{align}
&i\hbar\frac{\partial}{\partial t}\hat\rho(t)=[\hat H(t),\hat\rho(t)]+\mathcal{L}_D[\hat\rho(t)], \\
&\mathcal{L}_D[\hat\rho(t)]=i\sum_{k=1}^2\left(\hat L_k\hat\rho(t)\hat L_k^\dagger-\frac{1}{2}\{\hat L_k^\dagger\hat L_k,\hat\rho(t)\}\right). \notag
\end{align}
The master equation is built using the full Hamiltonian of the coupled qubits $\hat H$ as given by Eq. (\ref{eq:coupled-transmons}). However, in order to speed up the simulations, we work on the eigenbasis of the Hamiltonian at $t=0,$ truncating the basis to 60 states for both qubits \cite{SM}. Moreover, we express the Lindblad terms expressed in this basis as
\begin{equation}
\hat L_1=\!\!\sum_{n=-m}^{m}\sqrt{\frac{n+1}{T_1}}\ket{n}\bra{n+1},\ \hat L_2=\!\!\sum_{n=-m}^{m}\sqrt{\frac{n}{T_2^{*}}}\ket{n}\bra{n}.
\end{equation}
This generic model captures the particular setup specifications and source of errors through the characterics $T_1$ relaxation time and $T_2^{*}$ pure dephasing time.}

As shown in Fig.\ \ref{fig:lossy-cz}a, a decay time $T_1\sim \et{17}$ $\mu$s, such as those in experimentally available qubits\ \cite{Rol2019}, dominates the errors, equalizing all ramp methods. If we increase the quality of the qubits by an order of magnitude\ \cite{Place2020} see Fig.\ \ref{fig:lossy-cz}b, we find that the invariants and variational ramps become significantly better and more robust at long times, which contradicts the myth that staying close to the eigentstates leads to higher quality gates\ \cite{Barends2014}. \et{As show in Fig.\ \ref{fig:dephased-cz} a similar performance of the controls is found when dephasing is taken into account. A fast dephasing time $T_2^{*}=17$ $\mu$s equalizes the different designs, whereas the invariant and variational approaches show the robustness at $T_2^{*}=300$ $\mu$s.}

\jj{Finally, it is important to remark that some basis truncation is required for the simulations to converge in reasonable time and resources. However, we find that the choice of basis $\ket{n}$ is not very relevant: since the overlap between the low-energy sectors of the Hamiltonian at different times exceeds $99.99\%,$ different choices provide very similar plots of the fidelity. Moreover, we have checked convergence with respect to the truncation size of the basis, as well as with respect to the numerical integration methods.}

\subsection{Pulse bandwidth and distortions}
\label{sect:distortions}

\begin{figure}[t]
  \centering
  \includegraphics[width=\linewidth]{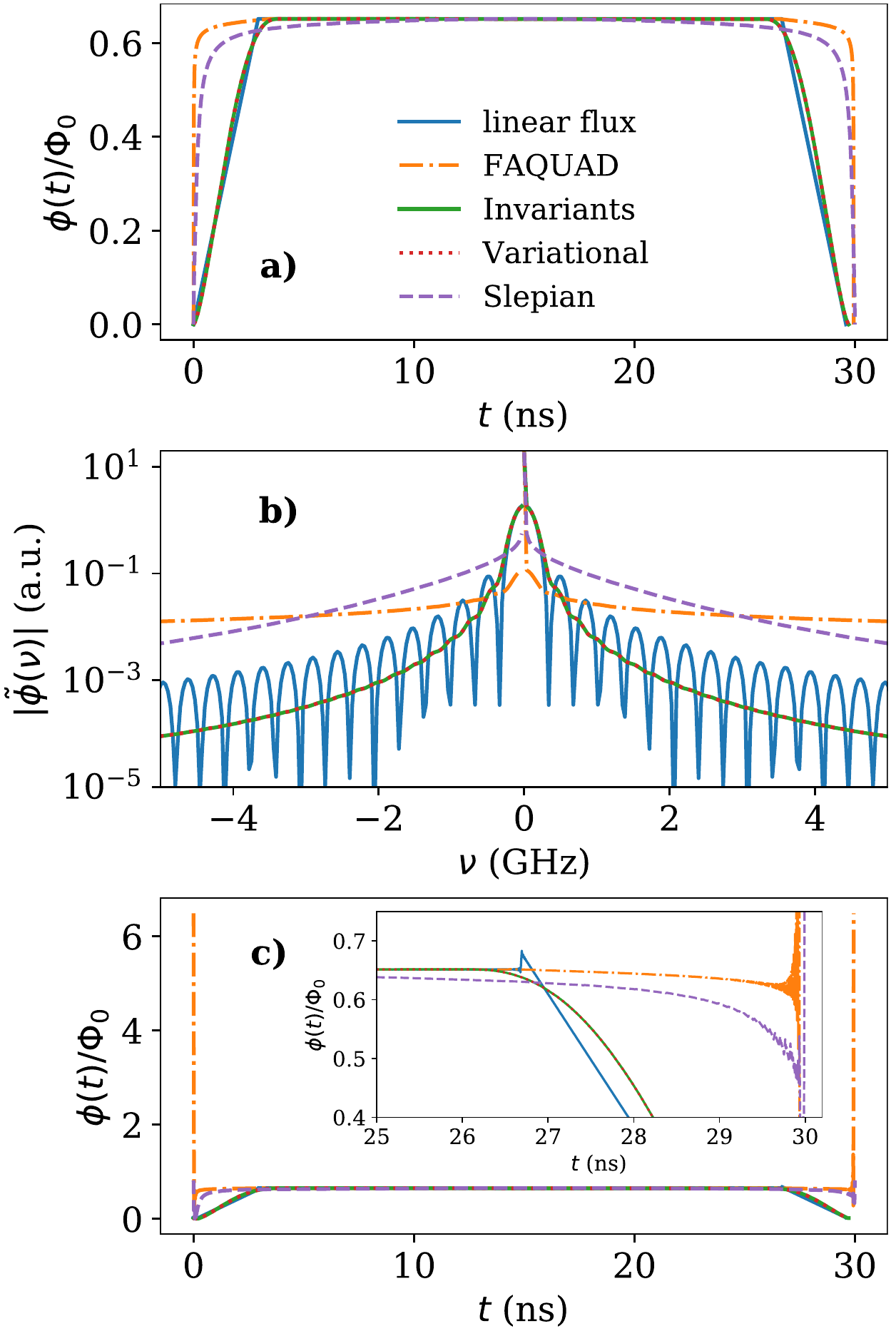}
  \caption{Optimized CZ gate for $T_\mathrm{gate}=30$ ns. (a) Flux pulse on the transmon qubit required to implement the gate. (b) Fourier transform of the pulse.
  \et{(c) Pre-distortioned flux pulse for a filter cut-off frequency $\omega_c=2\pi\times 200$ MHz. The inset shows a zoom-in of the ramp-down region.}}
  \label{fig:cz-pulses}
\end{figure}

\jj{So far we have studied idealized controls, which concern the electromagnetic fields that surround the qubit. However, in real experiments, those controls may suffer distortions due to the electrical response of the circuits that are used to generate and transport them. In particular, many superconducting quantum circuit experiments  have implicit or explicit low-pass filters that aim at reducing the noise around the qubits. The pulse $x(t)$ that must be injected to overcome those filters and generate a control $y(t)$ may be significantly different, involving sharp features and additional power.}

\jj{Given an input signal $x(t),$ the actual output $y(t)$ provided to the circuit is given by the intrinsic transfer function $h(t)$, through a convolution $y(t)=(h*x)(t)$. We can invert this equation in frequency domain. If $\mathcal{Y}(\omega)=\mathcal{F}[y(t)]$ is the Fourier transform of the desired control and $\mathcal{H}(\omega)=\mathcal{F}[h(t)]$ is the linear transfer function, the input signal that must be injected is given by $x(t)=\mathcal{F}^{-1}[\mathcal{Y}(\omega)/\mathcal{H}(\omega)],$ where $\mathcal{F}$ and $\mathcal{F}^{-1}$ denote the Fourier transform and its inverse, respectively.}

\et{We have considered a model transfer function $\mathcal{H}(\omega)=\frac{\omega_c}{\omega_c-i\omega}$ associated to a low-pass filter\ \cite{Langford2017} with cut-off frequency $\omega_c.$ From a visual inspection of the ideal controls, shown in Fig.\ \ref{fig:cz-pulses}a, we see that the FAQUAD and Slepian pulses are more likely to be affected by the low-pass filter, because they grow more rapidly and involve higher frequencies. This notion is reinforced by a Fourier transform of the controls. As seen in Fig.\ \ref{fig:cz-pulses}b,  the linear flux, invariant and variational pulses have the narowest bandwidth, requiring less precompensation. The calculation of the predistorted signals are depicted in Fig.\ \ref{fig:cz-pulses}c. We observe that the invariat and variational approaches are almost unmodified and fit the 200 MHz filter. The linear flux growth has a small distortion, and the FAQUAD and Slepian demand significant abrupt changes in the flux to produce the optimal controls. These abrupt changes will be hard to reproduce experimentally, leading to increased drive-induced dephasing, and additional compensation mechanisms to concatenate multiple gates\ \cite{Rol2019}.
}\\

\section{Summary}

In this work we have studied the implementation of a CZ gate using the avoided crossing between the $\ket{11}$ and $\ket{20}$ of two statically coupled transmons. We have shown that there are many different controls all of which lead to gates with excellent fidelities within times which approach the theoretical limit $\pi/J_2.$ For all controls, tuning the gate requires \textit{only} a calibration of the waiting time and of the ramp frequency.

Given the great variety of possible controls, are all choices created equal? We have argued that this is not the case. Some of these protocols, such as the invariants and variational methods, have a better performance due to their optimal control of leakage outside the computational basis, and a greater robustness against decoherence. Moreover, when we consider the physical parameters that are controlled---i.e. when we study the variation of flux that they demand---, we find that precisely those controls are the ones that have better properties of finite bandwidth and resilience to discretization [cf. Fig.\ \ref{fig:cz-pulses}].

\acknowledgments

Authors acknowledge N. K. Langford for fruitful discussions. J.J.G.-R. and  E.T. acknowledge support from Project PGC2018-094792-B-I00 (MCIU/AEI/FEDER,UE), CSIC Research Platform PTI-001, and CAM/FEDER Project No. S2018/TCS-4342 (QUITEMAD-CM).

\bibliographystyle{apsrev4-1}
\bibliography{main.bbl}

\end{document}